# Realization of La$_2$MnVO$_6$: Search for half-metallic antiferromagnetism?

J. Androulakis,$^{a,b}$ N. Katsarakis$^a$ and J. Giapintzakis$^{a*}$

$^a$*Institute of Electronic Structure and Laser, Foundation for Research and Technology – Hellas, PO Box 1527, Vasilika Vouton, 711 10 Heraklion, Crete, Greece*

$^b$*Department of Chemistry, University of Crete, Leoforos Knossou, 714 09 Heraklion, Crete, Greece*

Single-phase polycrystalline La$_2$MnVO$_6$ samples were synthesized by arc melting and characterized by X-ray diffraction, magnetization and resistivity measurements. We find that the compound has cubic (space group $Fm\overline{3}m$), partly ordered double perovskite structure. The sample exhibits ferrimagnetic behavior and variable-range hopping conductivity. We conclude based on the magnetic properties that both Mn and V ions are trivalent; moreover, the Mn$^{3+}$ ions are in a high-spin state, which is the reason that the compound is not a half-metallic antiferromagnet.

---
$^*$ Corresponding author:
giapintz@iesl.forth.gr
Fax: +30-810-391305

## I. INTRODUCTION

In a conventional antiferromagnet the suppression of the net magnetic moment at temperatures below the Neel temperature emanates either from a symmetry relation between sites with equal and opposite spin or the occurrence of a spin density wave. In both cases the electronic structure of the two spin directions is identical and thus, no polarization of the conduction electrons is possible. In the mid-nineties a seminal paper of H. van Leuken and R. A. de Groot,[1] suggested the possibility of realizing antiferromagnets of a truly new nature known as half-metallic antiferromagnets (HM-AFM). These systems should exhibit 100% spin polarization of the conduction electrons without showing a net magnetization. In a half-metallic material the exact cancellation of the local magnetic moments is provided by the requirement of the net moment to be integral and for certain materials this integer can be zero. The search for HM-AFM has been intensified due to the expectation that HM-AFM materials can be used in spin-polarized scanning tunneling microscopes as probes which will not perturb the spin character of the samples and in new spin electronic devices that rely on the spin polarization of the carriers.

The first proposed HM-AFM systems are CrMnSb and $V_7MnFe_8Sb_7In$ both required to have the Heusler $C1_b$ structure.[1] Due to the complexity of these materials a considerable effort has been spent to come up with other candidate materials, which may be easier to synthesize and thus search for HM-AFM characteristics, e.g., thiospinels[2] and ordered double perovskites (A'A"B'B"$O_6$; primes indicate the possibility of different ions).[3,4]

Pickett[3] has proposed, based on band calculations using the local spin-density approximation, that the double perovskite $La_2MnVO_6$ can be a promising candidate for exhibiting HM-AFM. The electronic structure calculations were based on the assumption that both Mn and V ions are trivalent and more importantly, a low spin state ($t_{2g}^4 e_g^0$; S=1) resulted for all $Mn^{3+}$ ions. The HM-AFM nature of $La_2MnVO_6$ has not been tested experimentally yet.

In this paper, we report on the structural, magnetic and transport properties of $La_2MnVO_6$ and discuss the theoretical prediction of HM-AFM behavior in the light of our results.

## II. EXPERIMENTAL PROCEDURE

The compound $La_2MnVO_6$ was prepared by arc melting a mixture of high purity oxides ($La_2O_3$: 99.99%, $Mn_2O_3$: 99%, $V_2O_3$: 99%) in argon atmosphere. The obtained polycrystalline ingots, which were placed in Ta boats and sealed in quartz tubes under a low-pressure reducing atmosphere, were annealed at ~700 ºC for several days. Several different batches were prepared to check the reproducibility of our data. Powder x-ray diffraction measurements were carried out on all synthesized samples using a Rigaku diffractometer with Cu $K_{\alpha 1}$ radiation to identify the crystallographic structure and the phase purity. Dc magnetization and ac susceptibility measurements were carried out on powder samples as a function of the applied magnetic field (up to 70 kOe) at temperatures between 5 and 360 K, using an extraction magnetometer. Electrical resistivity measurements were carried out on small rectangular bars using the four-probe method in the temperature range 160-300 K in a homemade apparatus.

## III. RESULTS AND DISCUSSION

The observed, calculated, and difference x-ray diffraction (XRD) profiles of the final compound are shown in Fig. 1. The CRYSFIRE suite[5] was used to index the profile and the FULLPROF program[6] to analyze the profile. $La_2MnVO_6$ is indexed on the basis of a cubic cell ($F m \overline{3} m$) with a=7.9073 A. One also notes that no impurity phase is detectable in the XRD, indicating a clean single phase.

The aforementioned results, i.e., the cubic cell and the doubling of the lattice parameters with respect to those for a random distribution of cations, indicate the formation of an ordered double-perovskite structure for this compound. This is an interesting observation since only one other compound, $Ba_2PrPtO_6$, is known to be an ordered double perovskite compound with zero charge difference between B' and B". In general, nearly ordered arrangements are encountered when the charge difference between B' and B" is two or more,[7] as for example in $Sr_2FeMoO_6$.

The temperature dependence of dc mass magnetization, M(T), measured with H=10 and 60 kOe is shown in Figs. 2a and 2b, respectively. For H=10 kOe and at temperatures below 100 K, we observe noticeable thermomagnetic irreversibility (TMI) between the zero-field-cooled (ZFC) and the field-cooled (FC) curves and the magnetization to increase steadily; in addition, the ZFC magnetization decreases abruptly below 7 K and exhibits a broad peak. On the other hand, for H=60 kOe there is no observable TMI and no broad peak. The temperature dependence of inverse molar magnetic susceptibility, $\chi^{-1}(T)$, for H=10 kOe is shown in Fig. 3. The shape of the $\chi^{-1}(T)$ curve suggests that the sample displays ferrimagnetic behavior. In the temperature range between 200 and 300 K, the $\chi^{-1}(T)$ data are fitted well with the Curie-Weiss law. The experimental value for the paramagnetic Curie temperature is $\Theta$=-162.1 K and the effective number of Bohr magnetons is $p_{eff}$=5.83 $\mu_B$. The high and negative value of $\Theta$ indicates that strong antiferromagnetic interactions are established among the magnetic ions at low temperatures. Also, the value of $p_{eff}$ suggests that the oxidation states of manganese and vanadium are $Mn^{3+}$: $3d^4$ (high spin: $t_{2g}^3\uparrow$, $t_{2g}^1\uparrow$; S=2) and $V^{3+}$: $3d^2$ ($t_{2g}^2\downarrow$; S=1) since the spin-only moment for $La_2Mn^{3+}V^{3+}O_6$ is calculated to be 5.65 $\mu_B$.

The dc magnetization as a function of an applied magnetic field measured at T=1.8 and 5K, is shown in Fig. 4. We observe a small hysteresis in the M-H curve, consistent with the ferrimagnetic behavior mentioned above. Because the magnetization remains unsaturated even under the highest magnetic field applied (H=60 kOe), we obtained the actual saturation magnetization $n_A$ by plotting M vs. 1/H and extrapolating the data to infinite field. We find that $n_A \approx 0.85$ $\mu_B$, which is considerably smaller than the expected value (2 $\mu_B$) for a long range ordered $Mn^{3+}$ - O - $V^{3+}$ double perovskite structure. This result suggests that there is only a partial ordering of $Mn^{3+}$ / $V^{3+}$ ions in the double perovskite $La_2MnVO_6$. The degree of the ordering of the B-cations needs to be further explored either by electron diffraction or Rietveld analysis of neutron powder diffraction profile.

The observation of TMI in the M(T) data shown in Fig. 2a, is one of the characteristics of spin-glassy behavior. In order to determine if this is indeed the case, we have measured the frequency dependence of the ac susceptibility. We find, as shown in Fig. 5, that there is no frequency dependence and thus, we conclude that the system does not exhibit glassy behavior. In view of the above, the observed downturn of the measured magnetic susceptibility below 25 K is certainly quite unexpected.

The temperature dependence of dc resistivity (Fig. 6) shows that the synthesized phase of $La_2MnVO_6$ is semiconducting. The lnR vs. $1/T^{1/4}$ plot is linear, indicating a variable range hopping (VRH) of charge carriers. The random potential fluctuations needed for VRH to apply can be attributed to the existence of domains exhibiting random $Mn^{3+}$-$V^{3+}$ arrangement.

The synthesized compound $La_2MnVO_6$ can be considered as 1:1 solid solution between $LaVO_3$ [Ref. 8] and $LaMnO_3$ [Ref. 9]. Both of these compounds are very similar and have the same structure, i.e., they have O′ orthorhombic structure, are insulators, and exhibit antiferromagnetic ordering with almost similar $T_N$ ($LaVO_3$: $T_N=140$ K and $LaMnO_3$: $T_N= 160$ K). Therefore, if the synthesized compound had a random sublattice, according to Vegard's law we would expect it to have an O′ orthorhombic structure and be an antiferromagnet with $T_N \sim 150$ K. The observed deviations from this law, i.e., the cubic structure with lattice parameter double of that of a conventional perovskite and the ferrimagnetic behavior, are of considerable interest and warrant of further investigation.

## IV. CONCLUSIONS

In conclusion, the primary goal of the study was to investigate the possibility of preparing the compound $La_2MnVO_6$ in such a way that Mn and V ions are trivalent and, moreover, Mn ions are in a low-spin state. As mentioned in the introduction, these conditions are required for $La_2MnVO_6$ to be HM-AFM material according to electronic structure calculations carried out by Pickett. Our results show that we have synthesized a single phase $La_2MnVO_6$ compound with a partly ordered double perovskite structure, which is not HM-AFM because, even though both Mn and V are trivalent, the Mn ions prefer to be in the high-spin state. However, the fact that we have successfully synthesized a partly ordered double perovskite with zero charge difference between the B-site ions is extremely interesting and needs further attention.

## V. Acknowledgements

We would like to thank Prof. Warren Pickett for his encouragement.

**FIGURE CAPTIONS**

FIG. 1. X-ray powder diffraction pattern of $La_2MnVO_6$ measured at 300 K (crosses). The solid line is the calculated profile and the vertical marks correspond to the position of the Bragg reflections. The difference is plotted at the bottom of the figure.

FIG. 2. Temperature dependence of mass magnetization of $La_2MnVO_6$ measured under a magnetic field of (a) 10 kOe and (b) 60 kOe.

FIG. 3. Temperature dependence of inverse magnetic molar susceptibility of $La_2MnVO_6$ measured under a magnetic field of 10 kOe. The line through the high temperature data points represents a fit to the Curie-Weiss law.

FIG. 4. Magnetic hysteresis loops of $La_2MnVO_6$ measured at T= 1.8 and 5 K.

FIG. 5. Temperature dependence of the real part of the ac susceptibility of $La_2MnVO_6$, for two frequencies of the ac driving field ($h_{ac}$ = 1 Oe) in $H_{dc}$ = 0.

FIG. 6. Temperature dependence of the resistance of $La_2MnVO_6$. Inset shows the lnR vs. $1/T^{1/4}$ plot for the same compound.

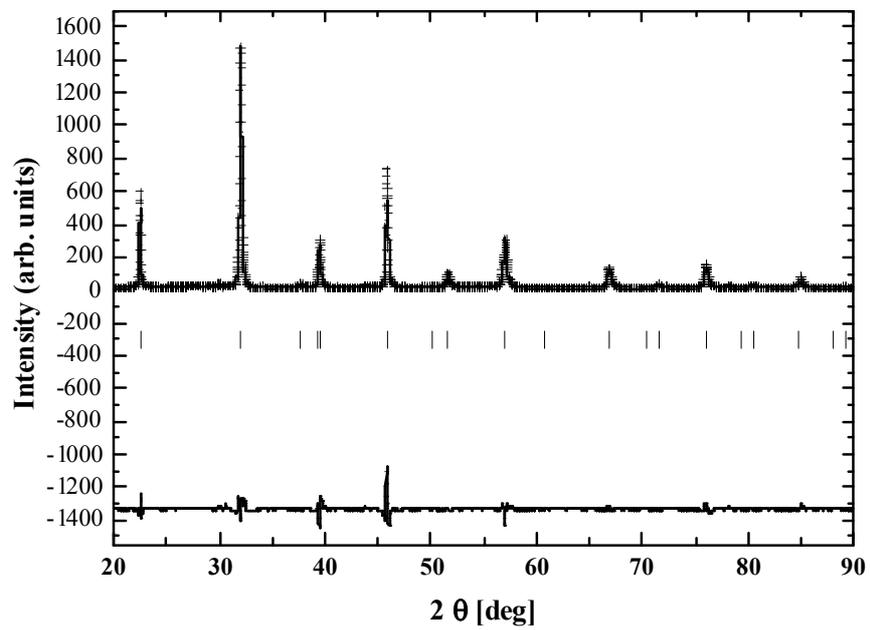

**Figure 1** .

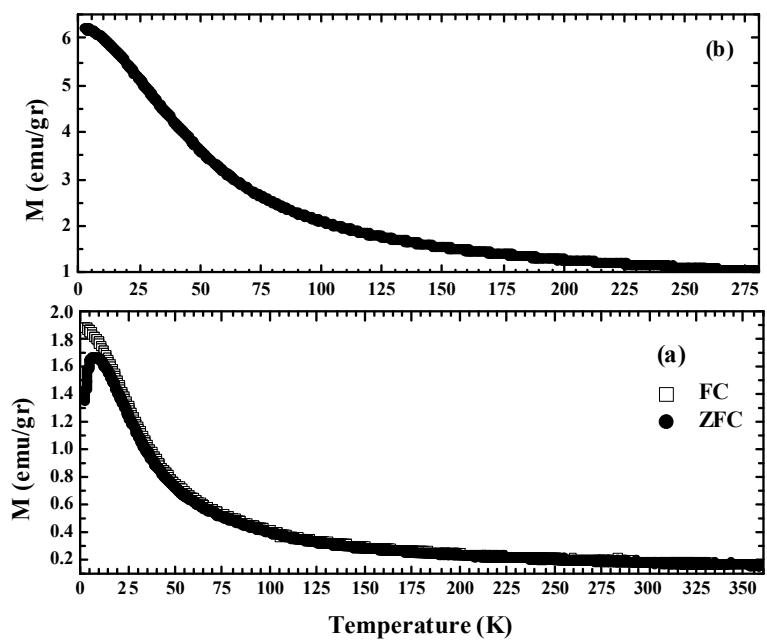

**Figure 2**.

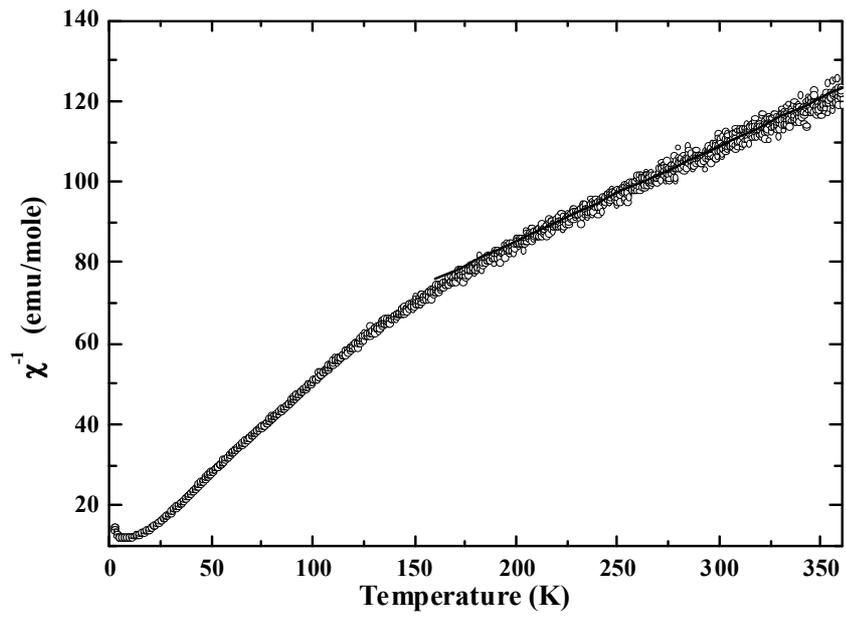

**Figure 3**.

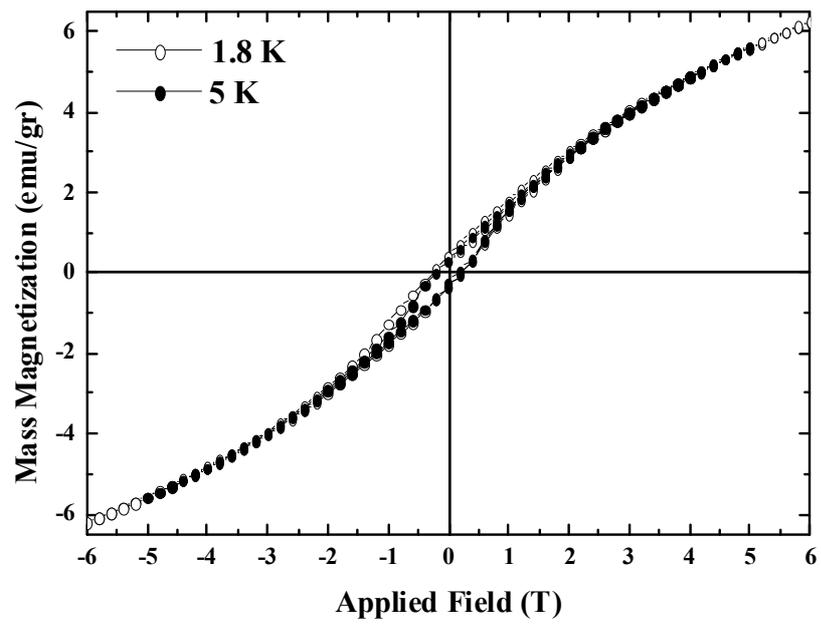

**Figure 4.**

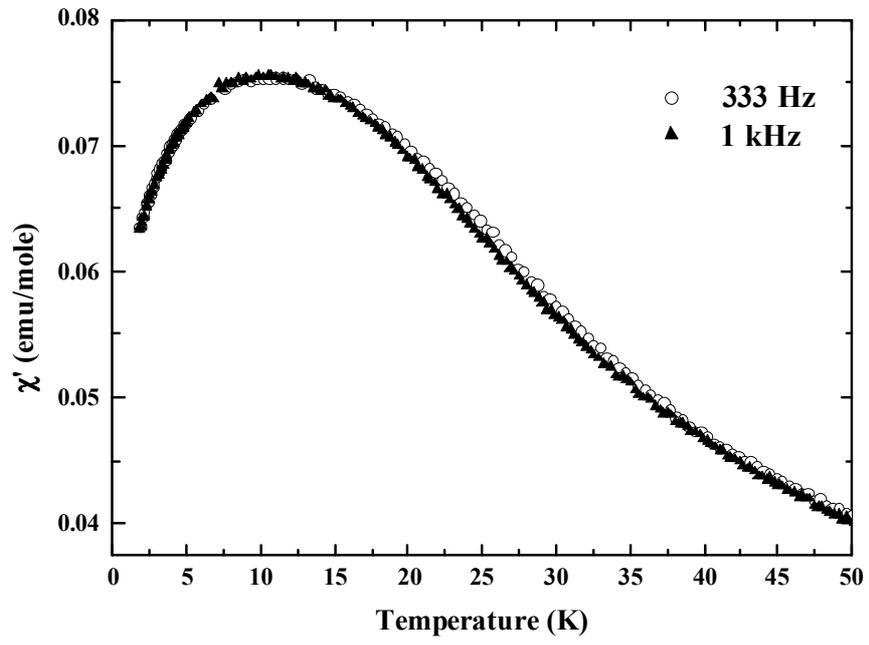

**Figure 5.**

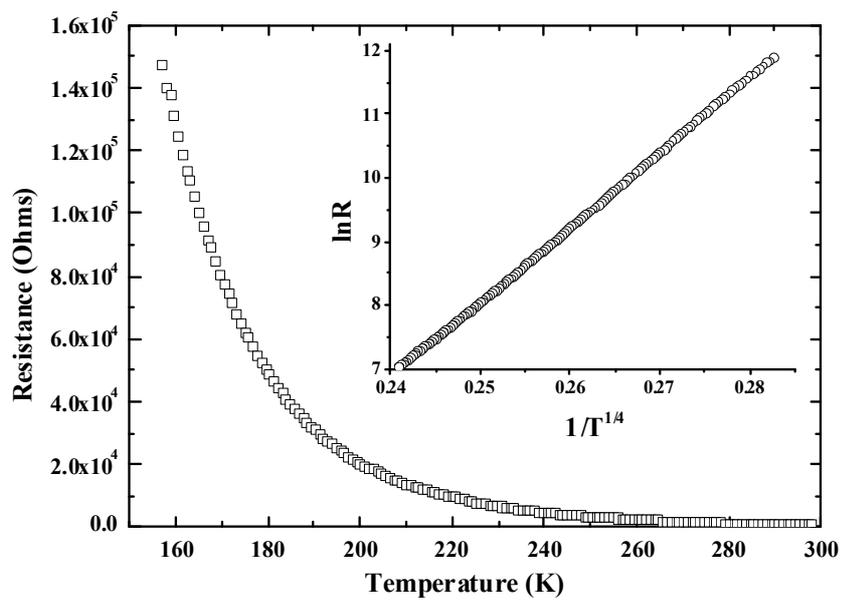

**Figure 6.**